# ZePoP: A Distributed Leader Election Protocol using the Delay-based Closeness Centrality for Peer-to-Peer Applications

Md Amjad Hossain
School of Business, Emporia State University
Emporia, KS, 66801
Email: mhossai1@emporia.edu

Javed I. Khan
Department of Computer Science, Kent State University
Kent, OH, 66801
Email: javed@cs.kent.edu

*Abstract*— **This paper presents ZePoP, a leader election protocol for distributed systems, optimizing a delay-based closeness centrality. We design the protocol specifically for the Peer to Peer(P2P) applications, where the leader peer (node) is responsible for collecting, processing, and redistributing data or control signals satisfying some timing constraints. The protocol elects an optimal leader node in the dynamically changing network and constructs a Data Collection and Distribution Tree (DCDT) rooted at the leader node. The elected optimal leader is closest to all nodes in the system compared to other nodes. We validate the proposed protocol through theoretical proofs as well as experimental results.**

## I. INTRODUCTION

Many distributed systems require a leader or coordinator node to synchronize system operations, collect data from all other nodes and redistribute the processed data as fast as possible. To ensure speedy operations, the leader must be quickly reachable from all other nodes in the network. Such a leader can be referred to as an optimal leader. In distributed applications on dynamic networks (i.e., P2P) the leader needs to be re-elected every time the leader loses its optimality. For example, in P2P video conferencing, a leader peer works as a Multipoint Control Unit (MCU) and must be elected dynamically as the network changes. The MCU must collect all streams as soon as possible, combine them into one stream, and return it to all participants[1, 2]. Similarly, in Federated Machine Learning on P2P systems, the dynamic leader must collect parameters of local models from the computing nodes and generate the global model. During the learning process, the leader also needs to calculate and distribute the model error among peers[3-7]. In these applications, an optimally positioned leader can speed up system operations significantly. The optimal leader can also be useful in many applications of Cyber-Physical Systems(CPS), such as the data collection and dissemination of control signals in wireless sensor networks, the swarm of robots, or the network of drones, IoT, etc.[8]. For the optimality, the election process must minimize the average distance to the leader from all other nodes or equivalently maximize the closeness centrality ( $C_x$ of node x), which is defined as [9],

$$C_x = \frac{n-1}{\sum_{y=1}^{n} D_{yx}} \quad (1)$$

Where $n$ is the total number of nodes in the network, and $D_{yx}$ is the distance from the node y to x. Earlier papers presented the leader election algorithms based on the known logical topologies of the systems such as ring, complete graph, tree, etc.[10-13]. However, the position-based optimal leader election must consider the end-to-end distance in the real network topology, which can be arbitrary in structure. Mega-Merger and Yo-yo[14] are among several universal leader election algorithms that work for arbitrary topology. But they elect the leader based on the unique identifiers of the nodes or their randomly proposed numbers. Many attempts have been made to merely estimate the closeness centrality using the neighborhood information of the nodes[15-17]. But only a few works have used the centrality measure in leader election, and a handful of them have used distributed algorithms. So far, the centrality measures considered are based on the hop count, i.e., $D_{yx}$ is the total hop count from node y to x [18]. But for delay-sensitive applications, a slow node or link might become a bottleneck, causing a loss in data communication. A delay-based closeness centrality in leader election can potentially find a better path by avoiding the extremely slow link.

So, in this paper, we present ZePoP (Zero Point Protocol), a distributed leader election protocol for the arbitrary network topology of P2P systems that elects the system leader by maximizing a delay-based closeness centrality. It defines a messaging scheme, election algorithm, node joining and departure algorithms. We validate the proposed protocol through theoretical proofs as well as experimental results. We show that the use of DCDT with an optimally placed root or leader can significantly improve the performance of applications. We organize the rest of the paper as follows: Section II discusses the related work. We formally define the leader election problem under consideration in section III. Section IV presents the complete ZePoP leader protocol for the dynamic network. Here, we discuss different algorithms and theoretical proof in detail. In section V, we show the experimental validation of the protocol. Section VI concludes the paper.

## II. RELATED WORK

Most of the leader election algorithms use unique identifiers of nodes for leader selection. Only a few uses closeness centrality (hop count based) for optimal leader election. Favier et al. in [18] presents a distributed eventual leader election algorithm for dynamic wireless networks using hop count based closeness centrality. In [19], W. Mary et al. presents a central algorithm for leader election using closeness centrality calculated from the adjacent matrix of the network. C. Kim and M. Wu present a distributed leader election mechanism with a tree-based centrality measure[20]. It forms a tree rooted at a random initiator of the algorithm. The centrality of each node is calculated using the layer and the depth information. The node with highest centrality is elected as the leader. They try to reduce the messages for the election but takes three phases. Moreover, the result highly depends on the initiator that start creating the tree. K. Mokhtarian and H. Jacobsen [21] discuss the algorithms for forming a minimum delay overlay multicast tree. The aim was to allow any node to build the tree if it requires to deliver delay-sensitive data to a group of receivers with minimum delay.

N. Andre et al. recently have worked on several centrality-based leader election algorithms[22]. In their opinion, "selecting a central node as the leader can significantly improve algorithm efficiency by reducing the network traffic or the time of convergence, especially in Large-scale lattice-based Modular Robots (LMRs) that form large-average-distance and large-diameter networks. In time-master-based synchronization protocols, placing the time-master at a central node leads to more synchronization precision in large-diameter networks as the precision of remote clock readings tends to decrease with the hop distance". Their ABC-Approximate-Center Election algorithm is like a leader election with hop-based closeness centrality[23]. They also propose the k-BFS SumSweep algorithm designed to elect an approximate center node[24]. Both algorithms are specially designed for choosing the center node as a leader in LMRs. They also worked on approximating the network centroid for large scale Embedded Systems[25]. For that, they use effective closeness centrality presented in[26] and tree-based leader election mechanism mentioned in[20]. However, all these algorithms elect the leader using hop-count based closeness centrality. They cannot optimize average delays from all nodes to the leader which is very important for delay-sensitive applications. Because a slow node or link in a path might become a bottleneck, increasing path delay and causing a loss in data communication. Moreover, these algorithms are not complete protocol to run the election in dynamic P2P environment.

A delay-based closeness centrality in leader election can potentially find a better path by avoiding the extremely slow link So, we propose a distributed leader election protocol that maximizes the delay-based closeness centrality (minimizes the average from all nodes to the leader). We show that the optimal placement of the leader in terms of delay can significantly improve the performance of the related P2P applications.

## III. ZePoP Protocol

### A. Problem Definition

Suppose there are *n* nodes in the dynamic distributed system (the nodes can come and go), where n ≥ 2. Assume that the graph G = (V, E) represents the P2P communication network where V is the set of all participant nodes, and *E* is the set of edges among them. One of the nodes, called leader, collects data from all other nodes, process data and distribute the processed or control data, among others. A Data Collection and Distribution Tree (DCDT) rooted at the leader node defines the datapath between nodes. So, we have to find a leader node $m \epsilon V$ that has the highest closeness centrality in G. In other words, if the closeness centrality of node *m* is $C_m$, then to select *m* as the leader, the leader election protocol must ensure $C_m \geq C_u, \forall_{u \in V,\ u \neq m}$ . The distance $D_{yx}$ as shown in (1) would be the path delay from *y* to *x*, which is the summation of point-to-point delays.

### B. Solution Overview

The ZePoP protocol defines a distributed leader election algorithm for the arbitrary network topology of P2P systems that can elect the system leader by maximizing a delay-based closeness centrality. We assume that nodes do not leave the network during the election process. It also defines all other P2P maintenance (supporting) algorithms for managing the dynamic properties of the network such as i) node joining, ii) node departure, and iii) detection of re-election. The election algorithm works in two phases. The first phase is to calculate the shortest path delays at each node from others as well as record the **branch weight information** for minimizing the number of leader candidates. In the second phase, each node determines its leader candidacy using the recorded branch weight. Then, each candidate calculates the closeness centrality and informs the value to others. All nodes in the network elect the candidate with the highest closeness centrality as the new leader. The direction information to the new leader is used to create the DCDT. In the subsequent sections, we discuss the messaging scheme, the algorithms, and proof of correctness.

| Fields | Node ID | d | D |
|---|---|---|---|
| Byte Offset | 0 ··· 1 | 2 ··· 5 | 6···9 |

(a) ELECTION, JOIN: the 'd' is used to carry point-to-point delay and 'D' is for carrying the path delay

| Fields | Node ID |
|---|---|
| Byte Offset | 0 ··· 1 |

(b) REPLY_ID, LEAVE, ARRIVAL, DEPART

| Fields | Node ID | Centrality(C) |
|---|---|---|
| Byte Offset | 0 ··· 1 | 2 ··· 5 |

(c) INFORM

Fig. 1 Message formats. In all messages, "Node ID" refers to the Source ID, but in REPLY_ID message, it is a New ID for the newly joined node.

### C. Message Scheme

The peers talk to each other by exchanging messages to elect the leader, form and reorganize the DCDT, inform node departure and arrival, etc. We assume that each node has a unique ID in the overlay network from S = {0, 1, 2, ..., $2^{16}$}. We form and maintain the P2P dynamic overlay network using the mechanism used in popular, fully distributed system Gnutella[27, 28]. Fig. 1 shows the structure of the messages used in the protocol. The REQUEST_ID and REPLY_ID are used for getting an ID for the newly joined node. PING and PONG messages are used to discover some existing nodes in the system, and their structures are the same as used in Gnutella. The structure of PING is also used for message REQUEST_ID. The JOIN message is used by the newly joined node to inform the leader about its joining. The ELECTION message is broadcasted by every node in the system to elect a new leader. The LEAVE message is used for informing the leader about node departure, and ARRIVAL is used by the current leader to inform all nodes about a node joining. A node broadcasts an INFORM message when it decides itself as a candidate for the leader. Besides, the nodes in the system exchange some strings with their direct neighbors, when necessary, with special prefix "VCONF".

*D. The election algorithm: Phase1*

**Table I: Notations and their initial values**

| Variables and descriptions | Initial values |
|---|---|
| $d_{xy}$ → link delay between x and y | $d_{xy}$ |
| $D_{sx}$ → path delay from node s to x | ∞ |
| $NB_x$ → set of neighbours of x in G | $NB_x$ |
| $\phi_s$ → A node sets this flag when it receives the first ELECTION message from the source s. | False |
| $\Psi_{sy}$ → The node x sets this flag if it forwards the ELECTION message of s to the neighbour y. | False |
| $\pitchfork_{sy}$ → $A$ node sets this flag if it receives the ELECTION message of node s via the neighbour y. | False |
| $O_{xy}$ → $the$ number of ELECTION messages forwarded by $x$ towards the neighbour $y$ in the shortest path. | 0 |
| $I_{xy}$ → number of ELECTION messages received by $x$ via neighbour $y$ in the shortest path ( $I_{xy} = O_{yx}$) | 0 |
| FG_LIST → The future parents in DCDT. | {} |
| CHILD_LIST→ The list of children in DCDT. | {} |

*a)* ***Phase1 overview:*** In the first phase of the algorithm, each node x broadcasts the ELECTION(E) message and calculates path delays $D_{sx}$ upon receipt of ELECTION messages from all others node $s \epsilon V, s \neq x$ . Each node continues processing ELECTION messages until it receives at least one message from all other nodes. Fig. 2 shows the phase1 of the election algorithm. Table I describes the notations used in the algorithm with their initial values that are assigned before each election. When an ELECTION message of source s travels from the node y to its neighbour x, it carries the cumulative link delays or the path delay $D_{sy}$ in the field D, as well as the link delay $d_{yx}$ in the field 'd'. Upon receipt of that message, the node x calculates $D'_{sx}$(line3) and use it with $\phi_s$to decide whether to drop the message or process further. If the received message is the first copy from s, it sets $\phi_s$ , increment message count, calculates $D_{sx}$, $\pitchfork_{sy}$ to **true**, and increment $I_{xy}$ by 1. It also forwards the message to other neighbours z and increments $O_{xz}$ by 1 and sets $\Psi_{sz}$ to **true** (line 4-9). While processing the next copies of ELECTION message from s, each time the node x updates $I_{yx}$ and $O_{xy}$ considering the values $D_{sx}$, $D'_{sx}$ , $D'_{sy}$ , $\pitchfork_{sy}$ and $\Psi_{sy}$ (line 10-23). $I_{xy}$ and $O_{xy}$ are considered as the **branch weight information**. During the updates, x classifies the node s in one of the three categories, as discussed below. The node x also identifies the slow direct links with the neighbours so that it can avoid them (marked as **dead links**) during communication (line 25-27). On the dead links, both $I_{xy}$ and $O_{xy}$ would be 0.

*b)* ***Node Classification for Neighbourhood Comparison:*** As the node x receive election messages and updates $I_{xy}$ and $O_{xy}$ considering the values $D_{sx}$ , $D'_{sx}$ , $D'_{sy}$ , $\pitchfork_{sy}$ and $\Psi_{sy}$, it classifies each source node s with respect to the link (x,y) in one the classes A, B, or C as shown in Fig 3. The Node-Set A contains all participants, including x such that they have the shortest path to y only via x. Similarly, nodes in B, including y have the shortest-path to x only via y.

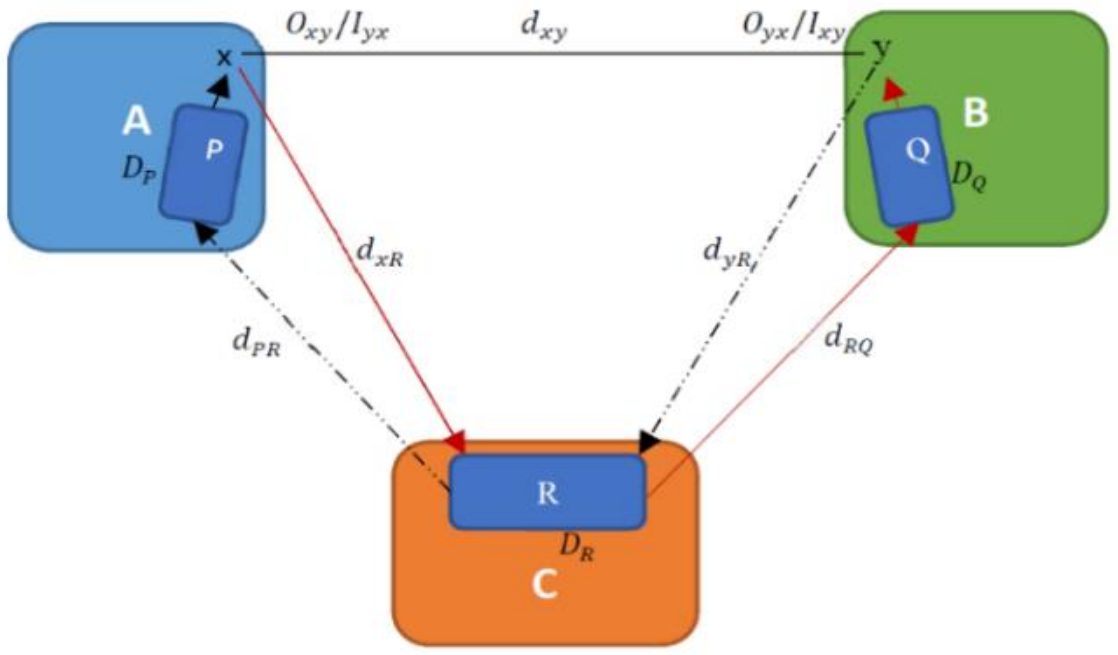

*Fig. 3. Node classification based on ELECTION messages.*

Participant nodes in C have the shortest path to x or y without going through y or x, respectively. P, Q, and R are the subset of nodes from A, B, and C, respectively, that the paths between x and y use.

1) If the network topology has no alternate path between *x* and *y* except the direct one, the classification is straight forward with C empty. All nodes reaching x via y are in set B, and the rest of the nodes are behind x, so they are in set A.

**2)** If there are multiple alternative paths between x and y then some nodes $R \subseteq C$ will be along the paths. The route from x to y might also include some nodes $Q \subseteq B$ and y to x might include some nodes $P \subseteq B$. As a node *x* receives ELECTION message from other nodes via these different paths, it ***classifies s with respect to link (x, y)***.

**ZePoP: Phase1**

**Recording (x, n)** // Executed at node x

**Begin**

0. message_count = 0; initialize the variables in Table I

Repeat step 1-23 until message_count<n-1

1. receive (E, y) // receives Election message via neighbour y
2. $D'_{sy} \leftarrow E.D$; s = E. NodeID
3. $D'_{sx} \leftarrow D'_{sy} + (E.d + d_{yx})/2$
4. **if** $\phi_s = \mathbf{false}$ // receiving a message from s for the first time
5. $\phi_s \leftarrow \mathbf{true}$;
6. Message count += 1]
7. $D_{sx} = D'_{sx}$
8. Accept (E, s, y)
9. Forward (E, s, y)
10. **else if** $D'_{sx} < D_{sx}$ // better message has arrived.
11. resetDirection(s)
12. $D_{sx} = D'_{sx}$
13. Accept (E, s, y)
14. adjustSend (s, y)
15. Forward (E, s, y)
16. **else if** $D'_{sx} = \mathbf{D_{sx}}$ // Equally better message received before
17. Accept (E, s, y)
18. adjustSend(s, y)
19. **else if** $D_{sx} + d_{xy} > D'_{sy}$ // x, y both are in equally better position for s. s in set C
20. $D_{sy} = D'_{sy}$
21. $T_x^C = T_x^C + D_{sx}$
22. $T_y^C = T_y^C + D'_{sy}$
23. adjustSend (s, y)
24. **else if s=x** // message coming back to x, direct x-> y is slow
25. If $(D'_{sy} < d_{xy})$ then $D_{sy} = D'_{sy}$; adjustSend (s, y)
26. **else if s = y and $\mathbf{D_{yx}} < \infty$ and $\mathbf{D_{yx}} < \mathbf{d_{yx}}$** // direct link y->x is slow
27. adjustSend (s, y)

**End**

**Note:** the variables with prime(') are temporary locals

**Methods**

adjustSend (s, y) // adjust # of sends to neighbor y

**Begin**

1. **if** $\Psi_{sy} \leftarrow \mathbf{true}$ **then**
2. $O_{xy} \leftarrow O_{xy} - 1$ ; $\Psi_{sy} \leftarrow \mathbf{false}$

**End**

Accept (E, s, y) // adjust # of receives from neighbour y

**Begin**

1. $G_s \leftarrow [G_s, y]$ // direction or gateway
2. **if** ($\pitchfork_{sy} = \mathbf{false}$) **then** // receipt flag of s via y is false
3. $\pitchfork_{sy} \leftarrow$ true ; $I_{xy} \leftarrow I_{xy} + 1$
4. $D_{sy} = E.D$

**End**

Forward (E, s, y) // better messages are forwarded

**Begin**

1. $E.D \leftarrow D_{sx}$ // update the message
2. **for** $z \in NB_x, z \neq y$
3. **If** $D_{sz} > (D_{sx} + d_{xz})$ **and** $\Psi_{sz} = \mathbf{false}$ **then**
4. $\Psi_{sz} \leftarrow \mathbf{true}$ // send flag
5. $O_{xz} \leftarrow O_{xz} + 1$
6. **If** $\Psi_{sz} \leftarrow$ **true then //** forwarded flag true
7. $D_{sz} \leftarrow D_{sx} + d_{xz}$ // update s to z path delay
8. $E.d \leftarrow d_{xz}$
9. Send (E, z)

**End**

ResetDirection(s) // the shortest path direct of node s

**Begin**

1. **for** $y \in G_s$
2. $I_{xy} \leftarrow I_{xy} - 1$
3. $\pitchfork_{sy} \leftarrow \mathbf{false}$ // reset receive the flag of s via y

**End**

ReceiveInform ($C'_x$, Leader, Leaderdirection)

**Begin**

1. Receive (I, g) // receive INFORM message
2. if I.C > $C'_x$ or I.C = $C'_x$ and Leader>I.nodeID **then**
3. Leader ← I.nodeID
4. $C'_x$ ← I.C
5. Leaderdirection ← g

**End**

Fig. 2. ZePoP: Phase1: Calculating the shortest path delays.

Finally, the x considers the node *s* in,

i. Set A, if $D_{sy} > D_{sx} + d_{xy}$, $D_{pq}$ *is* initialized $\infty$ . $D_{sy} = \infty$, if x has no copies of ELECTION message from s via y.

ii. Set B, if $D_{sx} > D_{sy} + d_{yx}$ or $D_{sy} < D_{sx} - d_{yx}$,

iii. Set C if $D_{sx} < D_{sy} + d_{yx}$ but $D_{sy} < D_{sx} + d_{xy}$ i.e. $(D_{sx} + d_{xy}) \leq D_{sy} \leq (D_{sx} + d_{xy})$

This classification is shown diagrammatically in Fig. 4. Thus, for each live link (x, y) the node x knows that $O_{xy} = |A|$ and $I_{xy} = |B|$. Now, suppose $T_x^C$ and $T_y^C$ are the total delay from the nodes in C, to node x and y, respectively. They are defined as follows, $T_x^C = \sum_{c\epsilon C} D_{cx}$ and $T_y^C = \sum_{c\epsilon C} D_{cy}$. The algorithm aims to enable each node *x* to decide if it is a leader candidate based on neighbourhood comparison of recorded values. For that, the node *x* also record both $T_x^C$ and $T_y^C$ as shown in lines 21 and 22. The algorithm uses the values of $O_{xy}$ , $I_{xy}$ , $T_x^C$ and $T_y^C$ to determine the leader candidates in phase2.

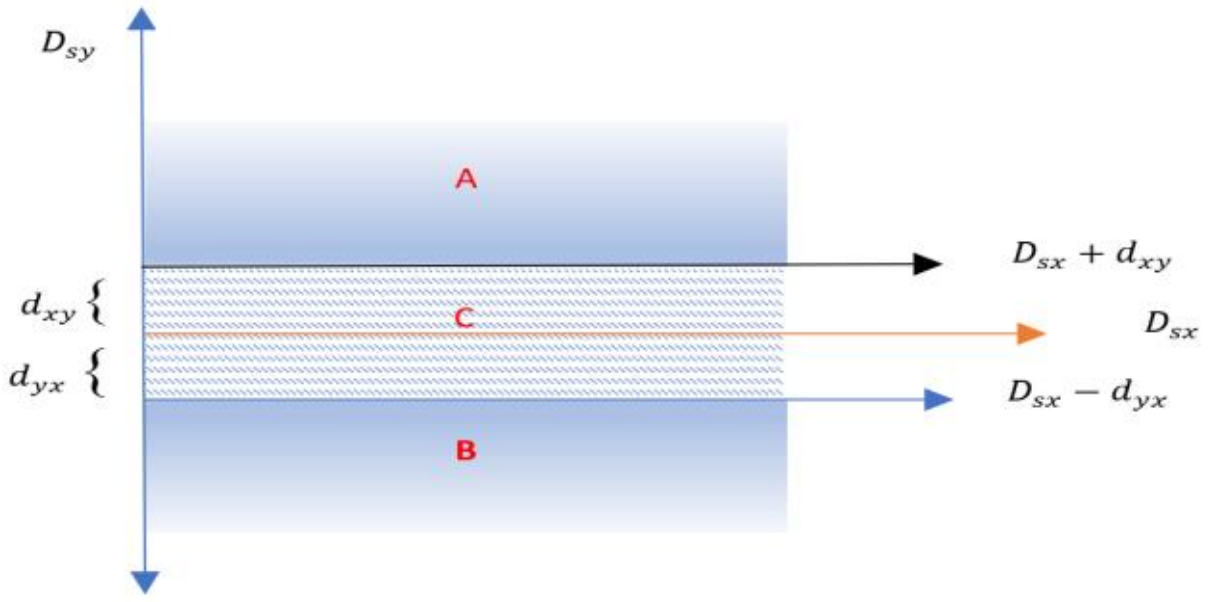

Fig. 4 Node classification with respect to link (x, y)

## E. *The election algorithm: Phase2*

In the second phase, the algorithm first aims to reduce the number of leader candidates by using the recorded values in phase1. Then, it elects one of the candidates as the new leader. Each node $x$ checks if it is a better candidate for the leader compared to its each neighbour y. For $x$ to be a better candidate, it must satisfy the following condition,

$$\text{Closeness Centrality, } C_x > C_y$$
$$\text{Or } \frac{|V|}{T_x} > \frac{|V|}{T_y}$$
$$\text{Or } T_x < T_y$$

Where the $T_x$ and $T_y$ are the total delay from all other nodes at x and its neighbour y respectively. We can calculate them as,

$$T_x = T_A + T_B + T_x^C + I_{xy} d_{yx}$$
$$T_y = T_A + T_B + T_y^C + O_{xy} d_{xy}$$

Where the $T_A$ and $T_B$ are the total delays from the nodes in set A and B, to the node x and y, respectively. So, assuming $d_{xy} = d_{yx}$ and $d_{xy} > 0$, $x$ must satisfy,

$$O_{xy} > \underbrace{I_{xy} + \frac{T_x^C - T_y^C}{d_{xy}}}_{I'_{xy}}$$

$$\text{So, } O_{xy} > I'_{xy} \qquad (2)$$

However, there is a possibility that the $d_{xy}$ would be zero. Then the condition becomes,

$$T_x^C < T_y^C \qquad (3)$$

Suppose the term $\delta_{xy}$ is $O_{xy} - I'_{xy}$, $if\ d_{xy} > 0$ or $T_y^C - T_x^C$, $if\ d_{xy} = 0$.

If the *superiority* of node x is $\phi_{xy}$ with respect to the neighbour y, then

$$\phi_{xy} = \begin{cases} \boldsymbol{true} & if\ \delta_{xy} > 0 \\ \boldsymbol{false} & if\ \delta_{xy} < 0 \\ \boldsymbol{false} & if\ \delta_{xy} = 0\ and\ (y < x\ \&x! = L)\ or\ y = L \end{cases} \qquad (4)$$

Where $L$ is the ID of the current leader. When $\delta_{xy} = 0$, both x and y are the equally better candidate, but the lowest ID or the existing leader breaks the tie.

Now, a node $x$ can declare itself as a leader candidate only if $\phi_{xy} = true$ for all neighbours $y \epsilon NB_x$, (lines 3-4). For the election, all the candidates declare their closeness centrality to others through the INFORM message. Each node selects the node as the leader, whose INFORM message contains the highest centrality. Fig. 5 shows the phase2 of the algorithm. At the end of the phase2, all nodes exchange SUBS or USUBS string among the direct neighbours to form the DCDT. Each node knows its current parent and the children in *CHILD_LIST*. The root of the DCDT is the optimally placed leader. Thus, the DCDT is a delay-balanced tree and ready for any delay-sensitive data communication. A node can avoid getting disconnected from DCDT by picking another node from *FG_LIST* as the parent in case the current parent leaves.

```
Phase2: Leader election(x)
1.  candidacy, first_message ← false
2.  if ϕxy = true, ∀y∈NBx then candidacy ← true
3.  if candidacy = true then
4.      Calculate the closeness centrality Cx as eq. (1)
5.      Broadcast INFORM containing Cx
6.      first_message ← true
7.  Leader ←x
8.  C'x ← 0; t ← 0,
9.  T ←k ∗ max (Dsx, ∀s∈V )// k is a constant
10. LeaderDirection ← x
11. while first_message = false
12.     first_message= Check_Inform_arrival()// non-blocking check
13.     If first_message=true
14.         ReceiveInform(C'x, Leader, Leaderdirection)
15. while t<T // runs for diameter
16.     flag =false
17.     flag= Check_Inform_arrival()
18.     If first_message=true
19.         ReceiveInform(C'x, Leader, Leaderdirection)
20. FG_LIST ←{NBx − Leaderdirection}
21. Send SUBS message to Leaderdirection
22. Send USUBS message to all in FG_LIST
23. CHILD_LIST ←{}
24. for c=1 to |NBx|
25.     Receive(US, c)
26.     If US.Type=SUBS then   CHILD_LIST.add(c)
```

Fig. 5. ZePoP: Phase2: Leader selection

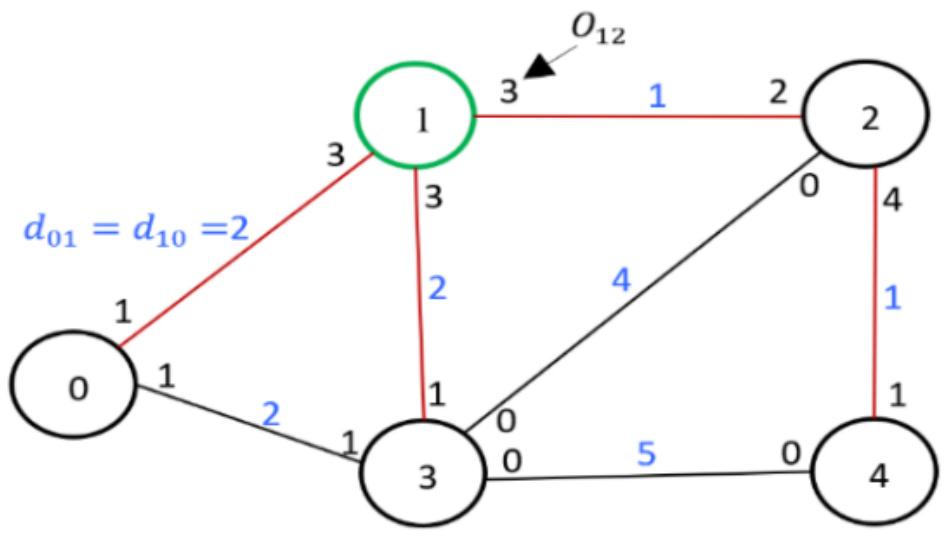

Fig 6. Example topology for leader election

An example of the leader election is shown in Fig. 6. The blue numbers on the edges are link delays. The black numbers represent the number of ELECTION messages forwarded in the shortest path, in the phase1 of the algorithm. For example, $O_{1,2} = 3$ means 3 ELECTION messages (for nodes 0, 1, 3) to node 2 will find the shortest path via node 1. In phase2, only node 1 will see $\phi_{1,j} = \boldsymbol{true}$ for all its neighbors $j$. So, node 1 is the only candidate for leader, and eventually, others elect it. According to the equation (1), the closeness centrality at the leader candidate 1 is **0.714**. The closeness centrality at nodes 0, 2, 3, and 4 are 0.45, 0.62, 0.45 and 0.45 respectively.

## F. *Minimum-Cost Spanning Tree (MST) vs. DCDT*

For leader election in the weighted network, where weight is the speed of the communication link, a very commonly used solution is: (i) create an MST (ii) pick any node or the center of the tree as the leader. However, MST cannot guarantee the highest closeness centrality. Let's consider the example in Fig. 6. The tree shown using the red lines is the DCDT, as well as an MST. However, if node 3 picks the link (3,0) instead of (3,1), the tree is still an MST but not the optimal DCDT. In that case, node 1 is still the center of the MST, but the closeness centrality would be 0.55, which is much lower than closeness centrality 0.714 of DCDT.

### G. *Maintaining Balanced DCDT*

The system with fixed topology initiates a new leader election only when the existing one fails. But, for the P2P applications in our consideration, the network must continuously respond to its dynamic nature and maintain the balance in DCDT. When a node joins or leaves the network, the network informs the current leader. The node joining and departure algorithms can be found in [1] .The leader *i* checks for violation of leader candidacy condition and it would start the election if for any neighbor *j*, $\phi_{i,j}$ becomes ***false***. For that, the leader keeps updating related parameters as it receives JOIN and LEAVE messages. If the current leader finds that an election is not required, then it only informs about the node arrival or departure to all nodes by broadcasting ARRIVAL or DEPART message, respectively. Thus, all peers can update their local variables, such as incrementing *n*.

Theorem 4.1 *The election protocol always finds a leader node maximizing the delay-based closeness centrality.*

*Proof:* For proof, first, we derive all possible network scenarios from Fig.3 and show that each scenario has at least one leader candidate.

*Base Case1- No cycle:* class C is empty. There is only one node x in A and one node y in class B. So, only two nodes in the system, Fig. 7 (a1). Both nodes are equally better candidates for leader, the lowest ID breaks the tie. But if we have a neighbour of y in B, then y is a better candidate than x, for any positive value of $d_{yk}$ Fig.7 (a2).

*Base Case2- Cyclic connectivity:* A single node from each class. Because of random delays on the links, we can have few possible subcases. For every pair of nodes x, y, (i) if $d_{xy} < (d_{xz} + d_{zy})$, where z is the third node in the cycle, then all nodes are equally better candidates, if $d_{xy} = d_{xz} = d_{zy}$ because $\phi_{i,j} = true$ for each i,j pair, Fig.7 (b1). But if $d_{zy} > d_{xz}$ and $d_{xz} \geq d_{xy}$ then only x is the leader candidate as both $\phi_{x,y} = true$ and $\phi_{x,z} = true$. (ii)But if $d_{xy} > (d_{xz} + d_{zy})$, then the direct link between x and y is considered as a **dead link,** Fig.7 (b2). On dead link (x, y), $O_{xy} = I_{xx} = 0$. In this case as well as when $d_{xy} = (d_{xz} + d_{zy})$, node *z* will be the leader candidate.

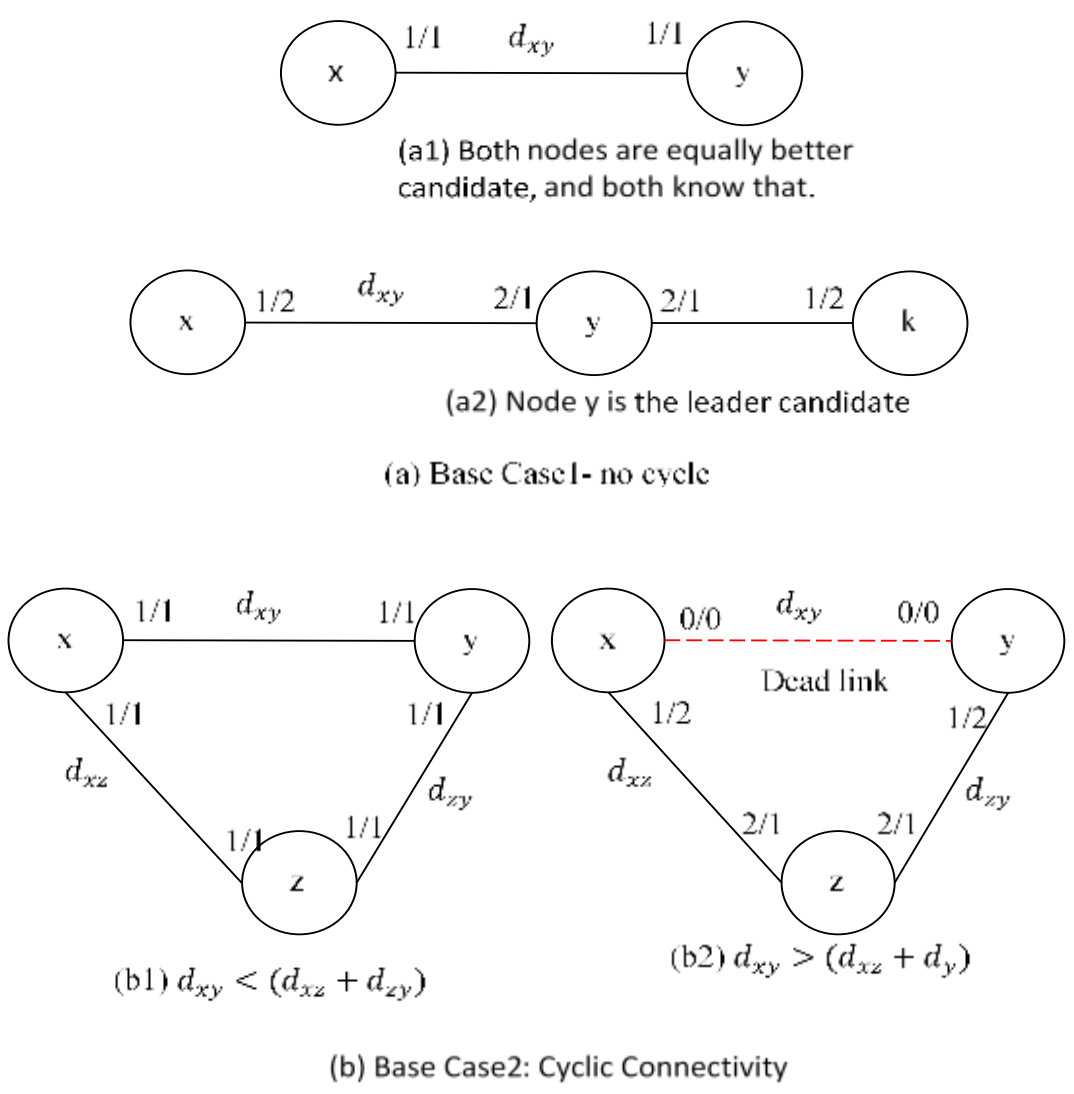

Fig. 7 Topology Base Cases and leader candidacy

*General case1- No cycle:* The topology is a tree structure. So, class C is empty for any link (x, y). We have one or more other nodes in A and B. In the tree structure, there is a single-center if it is a centered tree or two adjacent centers for bicentered tree. For single-center x, $\phi_{x,v} = true$ will be true for all neighbour v. So, x is the leader candidate. For the bicentered tree, two centers x and y are neighbours and $O_{xy} = I_{xy}$ is true. So, both x and y can be the candidates, but the algorithm chooses the lower node ID.

*General Case2 -Cyclic Connectivity:* There are multiple paths between x and y, including the direct link Fig. 3. The alternate paths can take 0 to many nodes from P, Q but at least one from R. Then we can locate the leader candidates as follow:

- if $O_{xy} = I_{xy}$ and $T_x^C = T_y^C$ then both set A and B are in the equal position to contain the leader candidate only if for any node $z \epsilon C$, $\phi_{xz} = true$ and $\phi_{yz} = true$. The set C also would have a candidate if $\phi_{zp} = true$ as well as $\phi_{zq} = true$ for $p \epsilon A$ and $q \epsilon B$. Fig 7-b2 is a special case for that.
- Now, if $\phi_{xy} = true$ and for any node $z \epsilon C$ $\phi_{xz} = true$ then the leader candidate is in set A. Similarly, it is possible to find any set B or C where the leader candidate(s) would exist.

After identifying the set for candidacy, we can move within the set towards the link say (x, q) where $\phi_{xq} = false$ until such a q exist. $\phi_{xq} = false$ means q is in a better position than x. This movement cannot be infinite if there is no message drop, i.e., we never complete a cycle. In other words, we can never have a general Case2 where y is better than x ($\phi_{xy} = false$), z is better than y ($\phi_{yz} = false$) and x is better than z ($\phi_{zx} = false$). Here $x \epsilon A, y \epsilon B$ $and$ $z \epsilon C$.

Eventually, we must reach an equilibrium position like the cases mentioned above where we have at least one leader candidate, and the candidate node, say x, sees for each neighbour j, $\phi_{xj} = true$. So, the algorithm always finds a very few but at least one leader candidate. As each candidate x ensures $\phi_{xj} = true$ for all neighbours $j \epsilon NB_x$, they are best in terms of closeness centrality compared to their neighbours. When each candidate informs its closeness centrality to others via INFORM message, they always elect the candidate whose INFORM message contains the highest closeness centrality.

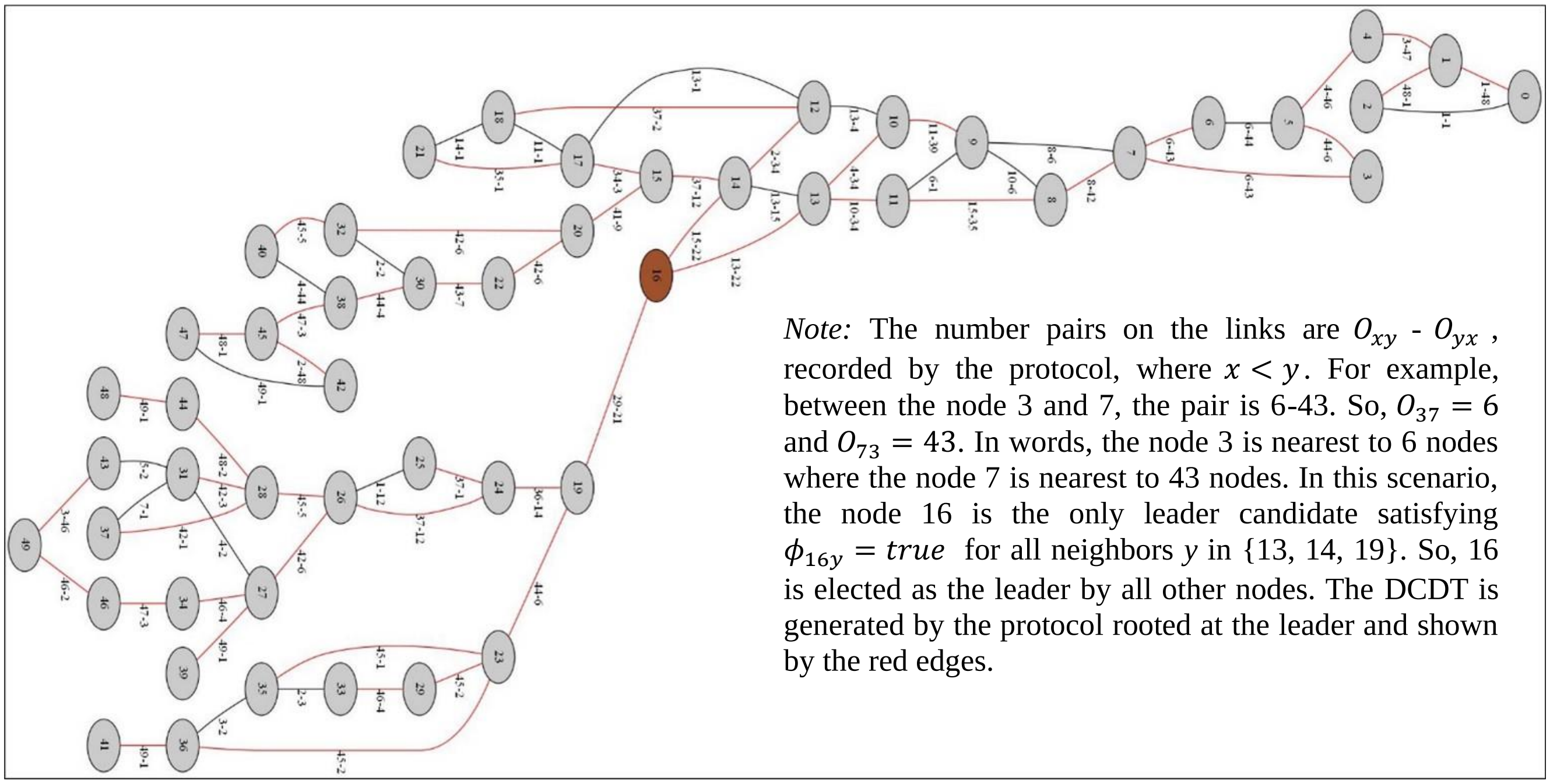

*Note:* The number pairs on the links are $O_{xy}$ - $O_{yx}$ , recorded by the protocol, where $x < y$. For example, between the node 3 and 7, the pair is 6-43. So, $O_{37} = 6$ and $O_{73} = 43$. In words, the node 3 is nearest to 6 nodes where the node 7 is nearest to 43 nodes. In this scenario, the node 16 is the only leader candidate satisfying $\phi_{16y} = true$ for all neighbors $y$ in {13, 14, 19}. So, 16 is elected as the leader by all other nodes. The DCDT is generated by the protocol rooted at the leader and shown by the red edges.

Fig 8. A randomly overlay network generated to emulate the ZePoP protocol

## IV. Experimental Results and Discussion

### A. *Protocol Implementation and Execution*

For experimental validation, we implement the protocol as a distributed application using MPI and C++. For running the application, we randomly generate P2P overlay networks on the local cluster nodes. We estimate the link delays in the overlay by 3-way message communication. We use these delays directly in the leader election algorithm. We run the protocol for different random overlay networks. Fig. 8 shows one of these networks. It has 50 nodes, i.e., n=50. On this network, node 16 is the only leader candidate according to the ZePoP, and eventually, all nodes elect it as the leader. The protocol also creates the DCDT (red edges) rooted at the optimally placed leader.

### B. *The delay-based closeness centrality comparison*

Fig. 9 shows the closeness centrality of each node in the network topology. The closeness centrality at the elected leader 16 is greater or equal to the centrality at all other nodes. Thus, the elected leader minimizes the average branch delay from all the nodes to the DCDT root (leader).

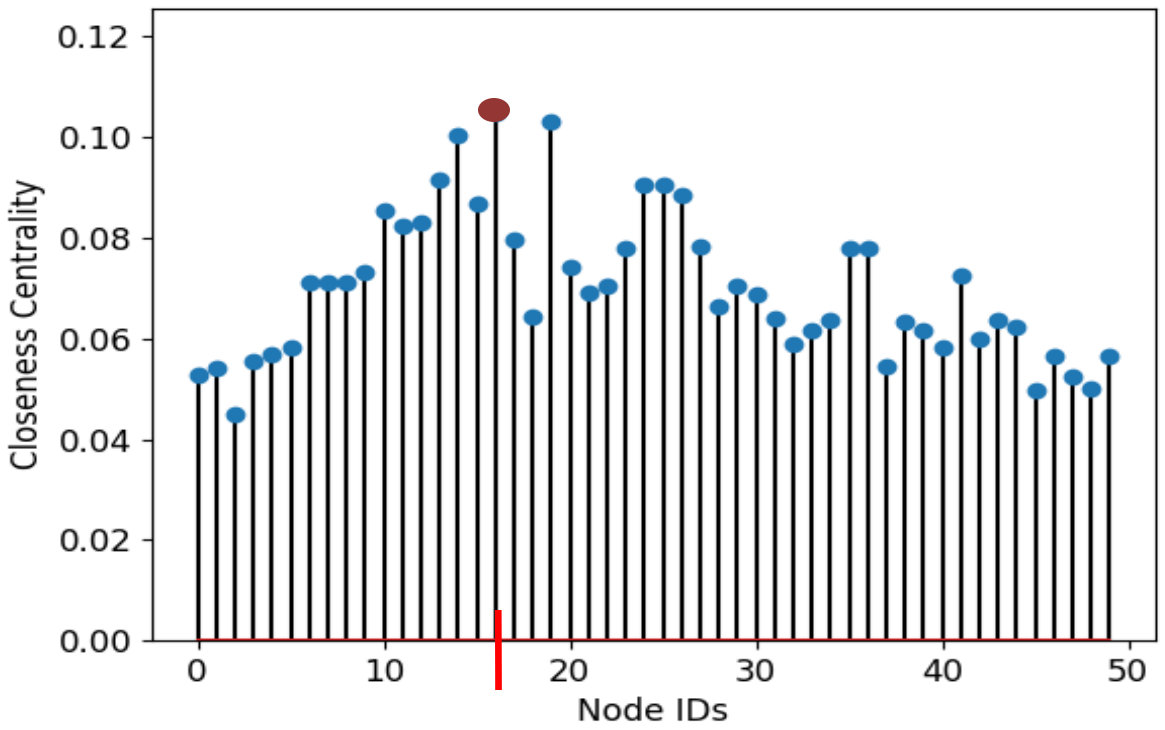

Fig. 9 Closeness Centrality comparison among all nodes

### C. *The Eccentricity Comparison*

We also compare the eccentricity of the nodes, which is the maximum branch delay at each node *(at node x, it is* $max(D_{s,x}), s \epsilon V$*)*. In a delay sensitive application if a node's data delivery time is higher than a set threshold, its data will be ignored. So, a leader with lower eccentricity will be more inclusive in data collections. The Fig. 10 shows that the eccentricity is minimum at node 16 as well as 13. The eccentricity at the leader will be near the minimum but not guaranteed to be the minimum.

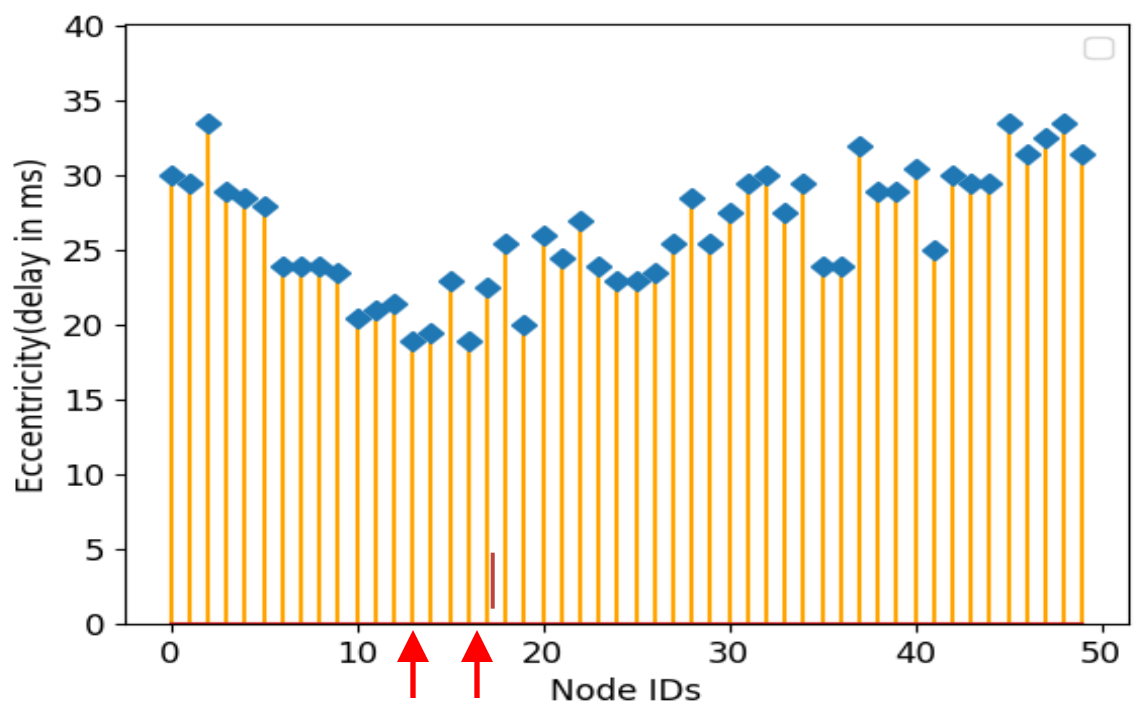

Fig. 10 comparison on eccentricity among all nodes

### D. Closeness Centrality vs Eccentricity

At each round of data collection and distribution, the leader must wait for a threshold which can be decided using closeness centrality or eccentricity. If the application needs data from all nodes in every round, then the eccentricity of the leader plus a fixed margin can be used as the waiting time. However, the slow links in the DCDT would have a significant impact on the eccentricity at each node, and the application might suffer from the scalability problem.

If the application can tolerate some data loss from some of the nodes, then we can use the closeness centrality or equivalently the average delay to speed up the data collection and distribution process by not waiting for the nodes behind the slow links. Suppose the shortest path delays from the nodes to the leader follows the normal distribution with mean $\mu$ and standard deviation $\sigma$. Then we can simply set waiting time

$$W = \mu + 3\sigma \quad (5)$$

where the $\mu$ is the mean or the average delay, and $\sigma$ is the standard deviation. The waiting time can ignore slow links or nodes and guarantee to receive data of almost 99% of the nodes. We can deduce some more useful schemas of waiting time management as shown in [29] s.

### E. Application: Video Conferencing

For experimental validation, we use the proposed ZePoP protocol in a P2P video-conferencing application where the elected leader works as a Multipoint Control Unit (MCU). We use MPI for messaging during the leader or the MCU election and socket programming with C++ for video data communication. For the experiment, we transmitted dummy video of frame size 1KB with header 23 bytes' header. The elected MCU generates a composite dummy video of the same size by taking a portion of everyone's video packet and returns to all nodes. The system is emulated on a local cluster with n = 50. The application starts with the node having ID 0 (assigned from MPI) and nodes *1, 2, ⋯, n-1* sequentially join the conference after every *15* seconds. As each new participant comes and connects to a random location of the network, the current MCU initiates the re-election to position the MCU optimally and maintain the balanced DCDT. We could also use the election initiation mechanism for determining the new leader or MCU election case. The randomly created final is the same as shown in Fig. 8.

For video conferencing, the location of the MCU is usually fixed. But for the P2P network, it is required to move the MCU dynamically to adapt to the dynamic nature of the network. To show the benefit of centrality based MCU placement, we consider node 0 as the static MCU. We compare average and maximum branch delays between the static MCU and the dynamic MCU (optimal leader) elected by our ZePoP protocol. Fig. 11 shows that both average and maximum branch delays are significantly lower for the dynamic MCU. Moreover, the average delay at the dynamic MCU is very low compared to the maximum delay(eccentricity). So, setting a fixed the waiting time based on the average delay i.e. only 10ms for this case would accumulate the video streams of almost all 50 participants at the dynamic MCU, whereas the static MCU will miss many participants' video.

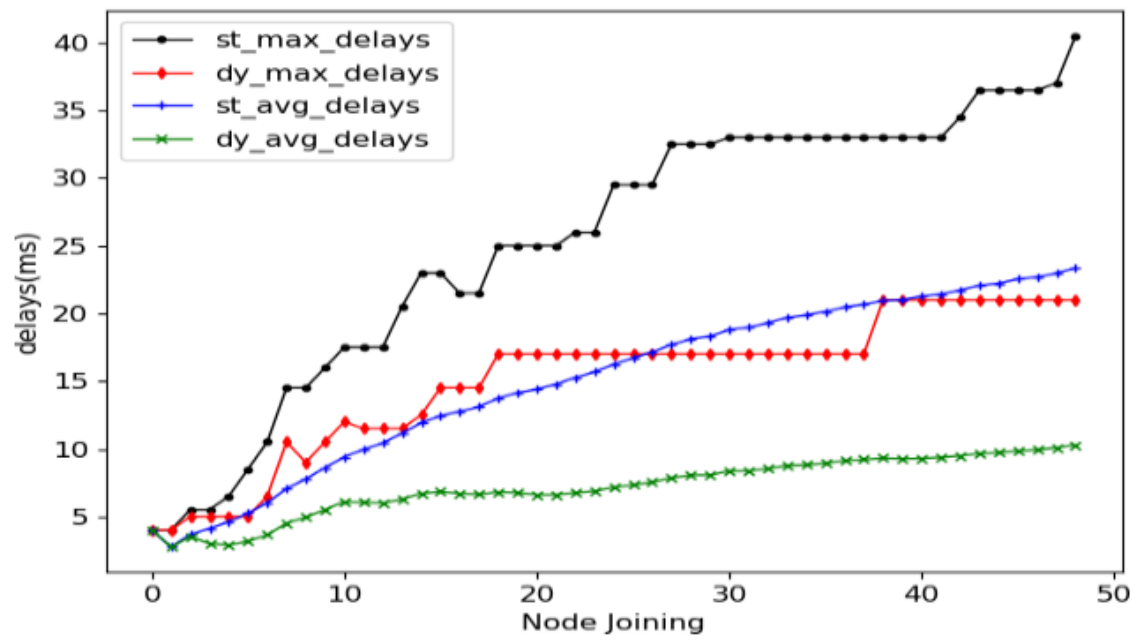

Fig. 11 Comparison on delays between the dynamic MCU and the static

### F. Message Complexity

In the first phase, each node broadcasts and always forwards the better ELECTION messages to the neighbours. However, we can use randomly filtered broadcasting to reduce some ELECTION messages in the network [30]. In our algorithm, each node requires only one copy of the ELECTION message from others. So, we can further minimize the ELECTION messages in the network by forwarding only one copy of the ELECTION message for each node. However, the message complexity would still be O(nE). In the future, we plan to improve the message complexity by allowing ELECTION messages to travel up to a certain distance in the network.

## V. CONCLUSION

In this paper, we present ZePoP, a distributed leader election protocol for distributed systems based on the delay-based closeness centrality. The election algorithm creates an optimal multicast tree called DCDT rooted at the optimally placed leader. The DCDT supports delay-constraint data collection and distribution applications. We also design multiple supporting algorithms so that the protocol can work on a dynamic P2P network as well. We show the validation of the proposed protocol both theoretically as well as with experimental results. We also demonstrate the benefit of the leader election protocol in a P2P video conferencing setting. We observe that the dynamic placement of a leader or MCU node can improve the system scalability as well as application performance by optimizing the average path delays. In the future, we would like to improve and analyze the protocol in terms of message complexities.